Bond Energy Sums in Benzene, Cyclohexatriene and Cyclohexane Prove Resonance Unnecessary

Raji Heyrovska

Institute of Biophysics, Academy of Sciences of the Czech Republic.

Email: rheyrovs@hotmail.com

**Abstract**

The recent new structure of benzene shows that it consists of three C atoms of radii as in graphite alternating with three C atoms with double bond radii. This is different from the hypothetical cyclohexatriene (Kekule structure) involving alternate double and single bonds. It was shown that the difference in the bond energy sum of the atomic structure of benzene from that of the Kekule structure is the energy (erroneously) interpreted as due to resonance. Here it is shown that the present structure of benzene also explains why its energy of hydrogenation into cyclohexane is different from that of cyclohexatriene.

**1. Introduction**

The structure of benzene has hitherto been assumed [1, 2] to be a combination of two Kekule (cyclohexatriene) structures in resonance. The difference between the observed and expected energy of hydrogenation of benzene (assuming it to be cyclohexatriene) to cyclohexane has been considered as the resonance energy [1] of benzene. It is explained briefly as follows [2a]: "The energy required to hydrogenate an isolated pi-bond is around 28.6 kcal/mol (120 kJ/mol). Thus, according to the VB picture of benzene (having three pi-bonds), the complete hydrogenation of benzene should require 85.8 kcal/mol (360 kJ/mol). However, the experimental heat of hydrogenation of benzene is around 49.8 kcal/mol (210 kJ/mol). The difference of 36 kcal/mol (150 kJ/mol) can be looked upon as a measure of resonance energy."

The present author has recently shown [3] that whereas the hypothetical [2b] cyclohexatriene (Kekule structure, see Fig. 1) involves three CC double bonds of length 1.34 Å alternating with three CC single bonds of length 1.54 Å, the atomic structure of benzene, as shown in Fig. 2a, consists of three carbon atoms ($C_{d.b.}$) with double bond radii ($R_{d.b.}=0.67$ Å) alternating with three atoms ($C_{r.b.}$) of the type in graphite (with delocalized charge) with the resonance bond radii ($R_{r.b.}=0.71$ Å). This new structure of benzene accounts for the observed equality of all six bond lengths (1.38 +/- 0.01 Å) and its bond energy sum of 1322 kcal/mole (observed value [1]: 1323 kcal/mole). The Kekule structure has a bond energy sum of 1286 kcal/mole [1], which is less by 36 kcal/mole than that of benzene, since it is structurally different from benzene. Therefore, the energy difference is not due to resonance between two (hypothetical) Kekule structures [1, 2].

## 2. Atomic structures and bond energy sums

The atomic structure of benzene [3] can be seen in Fig. 2a (the empty space at the center fits a circle with the double bond radius, $R_{d.b.}=0.67$ Å) and of cyclohexane in Fig. 2b (the central empty space can fit a circle with the single bond radius, $R_{s.b.}=0.77$ Å). Here it is shown that hydrogenations of the hypothetical cyclohexatriene (Fig. 1) and of benzene (Fig. 2a) with the consumption of $3H_2$ to cyclohexane (Fig. 2b) also differ by the same amount (36 kcal/mole) since benzene and cyclohexatriene themselves differ in energy by that amount [3] due to their different structures.

Using the bond energies in [1] for $C_{s.b.}C_{s.b.}$ single bonds (83 kcal/mole), $C_{d.b.}C_{d.b.}$ double bonds (146 kcal/mole) and CH aliphatic bonds (99.8 kcal/mole), in [3] for the $C_{d.b.}C_{r.b.}$ benzene bond (135.3 kcal/mole = (146 + 124.6)/2, where 124.6 kcal/mole is the graphitic resonance bond energy [4]) and in [4] for the CH benzyl bonds (85 kcal/mole), one obtains the following:

1) Cyclohexane (= 6 $C_{s.b.}C_{s.b.}$ single bonds + 12 CH aliphatic bonds): Bond energy sum = 83x6 + 99.8x12 = 1695.6 kcal/mole.

2) Cyclohexatriene (Kekule structure): 3 $C_{d.b.}C_{d.b.}$ double bonds + 3 $C_{s.b.}C_{s.b.}$ single bonds + 6 CH aliphatic bonds. The bond energy sum = (3x146 + 3x83 + 6x99.8) = 1286 kcal/mole, [1]. On adding the heat of hydrogenation (= 85.8 kcal/mole [2]) one obtains 1371.8 kcal/mole.

3) Benzene (= 6 $C_{d.b.}C_{r.b.}$ benzene bonds + 6 CH benzyl bonds): Bond energy sum = 6x135.3 + 6x85 = 1322 (= 1286 +36) kcal/mole (the observed value [1] is 1323 kcal/mole). The addition of the heat of hydrogenation (= 49.8 kcal/mole [2]), gives 1371.8 kcal/mole, as in 2).

The remainder, 1695.6 – 1371.8 = 323.8 = 3x108 kcal/mole ~ the energy of the covalent bonds in $3H_2$ used for hydrogenations in both cases 2) and 3).

Thus, the difference of 36 kcal/mole [1, 2] between the hydrogenation energies of benzene and cyclohexatriene is due to their structural difference alone. Therefore, the Kekule twin structures for benzene seem unnecessary. See [5] for the latest criticism of resonance.

**Fig. 1.** Cyclohexatriene (Kekule structure) assumed [1,2] for benzene, from [2].

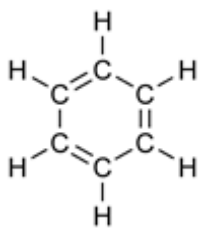

**Cyclohexatriene**

**(Kekule)**

**Fig. 2.** Atomic structures of a) benzene [3] and b) cyclohexane (this work). The CC bond lengths are (Å): $C_{d.b.} - C_{r.b.} = 0.67 + 0.71 = 1.38$ (benzene) and $C_{s.b.} - C_{s.b.} = 1.54$ (cyclohexane). The CH bond lengths are: $C_{d.b.}- H = 1.04$, $C_{r.b.}- H = 1.08$ and $C_{s.b.}- H = 1.14$.

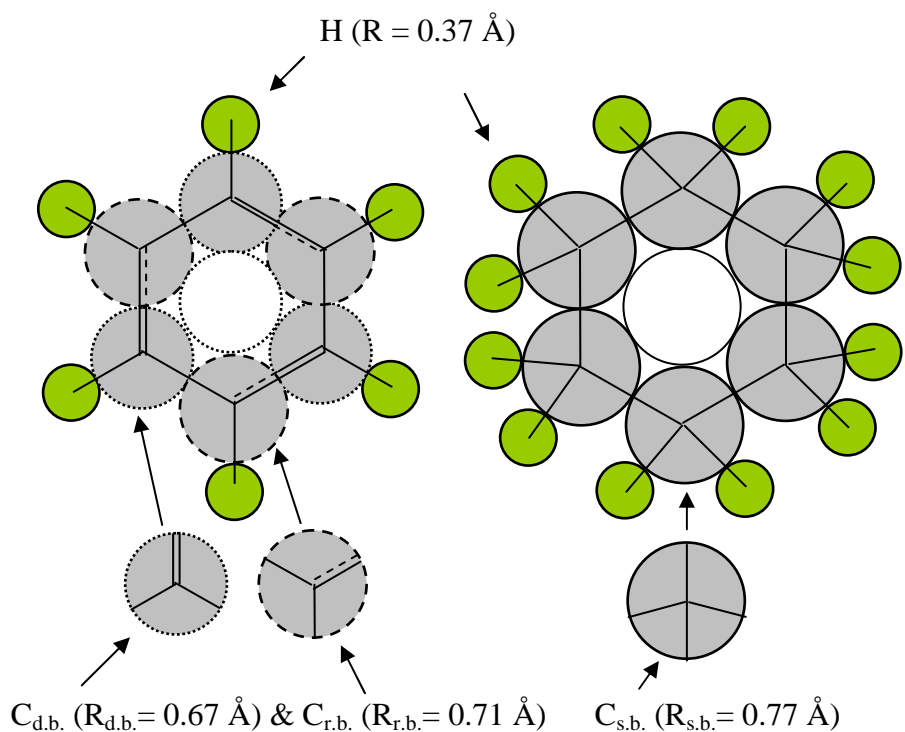

(a) Benzene  (b) Cyclohexane

H (R = 0.37 Å)

$C_{d.b.}$ ($R_{d.b.}= 0.67$ Å) & $C_{r.b.}$ ($R_{r.b.}= 0.71$ Å)   $C_{s.b.}$ ($R_{s.b.}= 0.77$ Å)